# Local aging effects in PuB$_4$: Growing inhomogeneity and slow dynamics of local-field fluctuations probed by $^{239}$Pu NMR

Seth B. Blackwell,[1,*] Riku Yamamoto,[1,*] Sean M. Thomas,[1]
Adam P. Dioguardi,[1] Samantha K. Cary,[1,2] Stosh A. Kozimor,[1]
Eric D. Bauer,[1] Filip Ronning,[1] and Michihiro Hirata[1,#]

[1]Los Alamos National Laboratory, Los Alamos, New Mexico 87545, USA
[2]Oak Ridge National Laboratory, 1 Bethel Valley Rd, Oak Ridge, Tennessee, 37830, USA
*These people have equally contributed to this work.
#mhirata@lanl.gov

The effect of self-irradiation damage can influence many properties of a radioactive material. Actinide materials involving the decay through alpha radiation have been frequently studied using techniques such as transport, thermodynamics, and x-ray diffraction. The use of nuclear magnetic resonance (NMR) spectroscopy to study such effects, however, has seen relatively little attention. Here, we use $^{239}$Pu NMR to study the local influence of self-damage in a single crystal of the candidate topological insulator plutonium tetraboride (PuB$_4$). We first characterize the anisotropy of the $^{239}$Pu resonance in a single crystal and confirm the local axial site symmetry inferred from previous polycrystalline measurements. Aging effects are then evaluated over the timeframe of six years. We find that, though the static NMR spectra may show a slight modulation in their shape, their field-rotation pattern reveals no change in Pu local site symmetry over time, suggesting that aging has a surprisingly small impact on the spatial distribution of the static hyperfine field. By contrast, aging has a prominent impact on the NMR relaxation processes and signal intensity. Specifically, aging-induced damage manifests itself as an increase in the spin-lattice relaxation time $T_1$, an increased distribution of $T_1$, and a signal intensity that decreases linearly by 20 % per year. The spin-spin relaxation time $T_2$ in the aged sample shows a strong variation across the spectrum as well as a drastic shortening towards lower temperature, suggesting growth of slow fluctuations of the hyperfine field that are linked to radiation-damage-induced inhomogeneity and could be responsible for the signal wipeout that develops over time.

## I. INTRODUCTION

$^{239}$Pu is the most widely studied isotope of plutonium, due to its availability and experimentally manageable half-life of 24,110 years. The decay of $^{239}$Pu nuclei involves both α decay and spontaneous fission, the former being the dominant decay mode. In calculations for metallic δ-Pu, such a decay process produces a 5.1-MeV α particle (helium nucleus) and an 88-keV $^{235}$U nucleus.[1] The α particle travels a rather long distance (an average of 11 μm) and loses 99.9 % of its kinetic energy through collisions with electrons, causing Pu displacements with approximately 265 Frenkel pairs (defects + interstitials). The $^{235}$U recoil nucleus travels a shorter distance (about 7.5 nm) and deposits 75% of its energy, producing 2,290 Frenkel pairs and causing nearly all the radiation damage in this process. After these decay events, most of these defects are thermally repaired but a remainder persists as defects in the lattice, which can affect material properties.[2]

Nuclear magnetic resonance (NMR) spectroscopy is a powerful local probe used across many scientific disciplines including solid-state physics, chemistry, and biology. In contrast to other spectroscopic methods, the small energy scale of the nuclear Zeeman energy probed by NMR provides high sensitivity to the low-energy electronic properties, making NMR an ideal tool to explore a material's magnetic and electronic properties in the vicinity of the NMR-active nuclei. In actinide science, NMR measurements on the ligand atoms have been used extensively to investigate local structures and dynamics, as has been reported for example in Ga stabilized $\delta$-Pu using $^{69,71}$Ga NMR.[3-7] Self-irradiation damage effects have been also evaluated in several materials, including $AmO_2$ by $^{17}$O NMR[8] and Pu containing zircons ($ZrSiO_4$) by magic-angle spinning $^{29}$Si NMR.[9] In the former, the $^{17}$O-NMR linewidth showed an appreciable broadening due to self-irradiation damage over a timeframe of a few weeks to years.

Direct NMR from the actinide atoms should have higher sensitivity to the 5$f$ electrons that mostly govern material's properties and has been, therefore, anticipated to be key to understanding the electronic and structural properties of actinide-based materials. Actinide NMR, however, has been limited so far due to the extremely large hyperfine coupling between unpaired $f$ electrons and nuclear spins. The magnetic hyperfine field at the nuclear positions can be very large (~ 100 T) and cause an excess broadening of the NMR response, spreading over orders of magnitude. Further, such pronounced local fields result in extremely fast nuclear spin-lattice relaxation times $T_1$ (<< 1 μs), making the detection of NMR signals particularly challenging. For uranium compounds, studies in the 1980's to 2000's detected $^{235}$U-NMR signals in a few exceptional cases where magnetic fluctuations of the otherwise unpaired $f$ electrons are absent, such as in the antiferromagnetic states of $^{235}$U-enriched $UO_2$[10,11] and $USb_2$[12] as well as in the liquid state of $UF_6$.[13] Regarding plutonium compounds, the signal detection proved even more difficult, and the discovery of a $^{239}$Pu-NMR signal in non-magnetic $PuO_2$ crystals was only reported in 2012.[14] Additional measurements in the same report further showed that variations in the oxygen content of $PuO_{2-\delta}$ caused a substantial reduction in the size of the NMR Knight shift by 5 – 15 %, indicating that $^{239}$Pu NMR is extremely sensitive to the local environment around Pu atoms. The second observation came in 2019, in the candidate topological insulator $PuB_4$ that has almost non-magnetic 5$f$-electron configurations.[15, 16]

In this work, we extend the $^{239}$Pu-NMR study from Ref. [16] with the same PuB$_4$ single crystal. We confirm that the local site symmetry for the Pu nucleus is effectively axially symmetric (i.e., locally tetragonal) from an angular rotation study of NMR spectra. We proceed with a systematic evaluation of the self-irradiation damage over a time span of six years. To date, detailed investigations on such damage effects using direct $^{239}$Pu NMR have not been reported. We find a relatively small influence of aging in the static spectrum but a notable change in the signal intensity and dynamic relaxation parameters. We show that these changes may be linked to a growing spatial distribution of dynamic hyperfine fields throughout the sample. Detailed analysis further reveals that a shortening of the spin-spin relaxation time $T_2$ develops due to growing slow fluctuations with decreasing temperature, which is linked to the signal wipeout observed as a function of aging. We also find that isochronal annealing at temperatures below 300 K can partially heal the radiation damage in a fresher sample condition in the first year, whereas it has a negligible impact once the sample accumulates significant damage after two years.

## II. MATERIALS AND METHODS

The aluminum-flux grown single crystal of PuB$_4$ used here (of about 17.4 mg and dimensions of 1×1×0.3 mm$^3$) is identical to the one used in Ref. [16], having the isotopic content of Pu (by weight %) of 93.96 % $^{239}$Pu, 5.908 % $^{240}$Pu, 0.098 % $^{241}$Pu, 0.025 % $^{242}$Pu, and 0.013 % $^{238}$Pu. The sample loaded in a copper NMR coil was encapsulated in a Stycast 1266 epoxy cube to avoid radioactive contamination. Titanium frits with 2-μm diameter pores were installed at both ends of coil's axis to seal the containment and to allow thermal contact with He exchange gas in our cryostat. Additional sample preparation details are given in the Supplementary Material of Ref. [16]. All the single crystal NMR measurements were performed between 90 and 1,800 days from the sample's initially synthesis. Results on the aged PuB$_4$ single crystal reported here are consistent with those reported earlier[16] on fresh powdered PuB$_4$ samples.

During six years of the measurement period, the single-crystal sample was thermally cycled several times between room temperature ($T$ = 300 K; for storage) and liquid $^4$He temperature ($T$ = 4 K; for measurements). See Fig. S1 for the full thermal history. Sample cooling and warming speeds were in the range ∼ 1–2 Kelvin/min to minimize thermal-contraction damage to the epoxy cube and to avoid potential contamination of the cryostat. All the NMR data to investigate aging effects were collected at a base temperature ∼ 4–5 K.

$^{239}$Pu-NMR measurements were performed using a commercial phase-coherent pulsed NMR spectrometer (Redstone, Tecmag Inc). The $^{239}$Pu nucleus has a nuclear spin $I$ = 1/2, which leads to a single NMR peak without any complication arising from the nuclear quadrupolar interaction relevant for the systems with $I \geq 1$. To obtain the NMR signal, standard spin-echo techniques were used at a fixed external magnetic field of 8.47–8.60 T with a fixed carrier frequency of 19.465 MHz. The field value was calibrated using the $^{63}$Cu-NMR signal of a copper coil wound around the crystal. Spin-echo signals were recorded at a fixed radio frequency after a conventional spin-echo ($t_{90°} - \tau - t_{180°}$) pulse sequence. Typical pulse widths were $t_{90°} = 8 - 9$ μs, which were sufficiently short to excite the full spectral bandwidth in our $^{239}$Pu spectra. Echo signals were then converted to NMR spectra using a Fast Fourier Transformation (FFT). The resonance frequencies in this work were converted to NMR Knight shifts using a bare gyromagnetic ratio $^{239}\gamma_n$ = 2π×2.29 MHz/T estimated in earlier studies of PuO$_2$.[14]

The nuclear spin-lattice relaxation time $T_1$ was determined from standard stretched-exponential fits to the recovery of nuclear magnetization after an inversion pulse sequence ($t_{180°} - t_{\text{wait}} - t_{90°} - \tau - t_{180°}$). In this case the time-dependent nuclear magnetization $M(t_{\text{wait}})$ can be expressed as

$$M(t_{\text{wait}}) = M(\infty)\left(1 - f \exp\left[-\left(\frac{t_{\text{wait}}}{T_1}\right)^{\beta}\right]\right), \quad (1)$$

where $M(\infty)$ is the nuclear magnetization at an equilibrium state, $f$ is the inversion fraction, and $\beta$ is the stretching exponent ($\beta \leq 1$) that measures the distribution of $T_1$ across the spectrum.

The spin-spin relaxation time $T_2$ was determined from the spin-echo decay of the magnetization $M(\tau)$ as a function of waiting time $\tau$ between two pulses. Because $T_2$ was strongly distributed across the NMR spectrum due to inhomogeneity and its size was orders of magnitude smaller than that of $T_1$, as we discuss below, the standard approach to decouple Gaussian and exponential contributions to the echo decay was not effective. Instead, a phenomenological stretched exponential form

$$M(\tau) = M_0 \exp\left[-\left(\frac{2\tau}{T_2}\right)^{\alpha}\right] \quad (2)$$

described the echo decay more appropriately. The exponent $\alpha$ ($1 \leq \alpha \leq 2$) accounts for the change in the character of the echo decay function at a different fraction of the spectrum.

To study the anisotropic nature of the $^{239}$Pu hyperfine interaction, we used a single-axis goniometer (NMR Service GmbH), with an angular resolution of 0.1 degrees, to measure the single-crystal $^{239}$Pu-NMR shift and relaxation rate as a function of rotation of the crystal with respect to an applied magnetic field.

### III. RESULTS

#### A. $^{239}$Pu-NMR spectrum and it's anisotropy in single crystal PuB$_4$

Figure 1(b) shows typical $^{239}$Pu-NMR spectra at $T$ = 10 K in a single crystal of PuB$_4$ taken after 90 days and 690 days from initial synthesis for the external magnetic field $B_0 (= \mu_0 H_0)$ of 8.5-8.6 T applied along the crystalline c-axis ($B_0 \| c$). (In-between these two measurements, the sample was thermally cycled several times between room temperature and 4 K as well as stored at 300 K for more than a year, as shown in Fig. S1.) A relatively broad single peak was assigned to the nuclear spin-level transition $I_z = 1/2 \leftrightarrow -1/2$ for the unique $^{239}$Pu site in the unit cell [see Fig. 1(a)]. Except for an overall intensity decrease by a factor of 1.66, about which we discuss below, the two $^{239}$Pu spectra exhibited almost identical line shapes, with no change in the linewidth and resonance frequency $f_0$ at the center of the peak, or equivalently in the value of the Knight shift $K$ defined as $f_0 = \gamma_n(1 + K)H_0$, where $\gamma_n$ is the nuclear gyromagnetic ratio. The NMR line profile reflects the histogram of the spatial distribution of the static local magnetic (hyperfine) field at the nuclear positions, and the linewidth is proportional to the Knight shift variation across the sample. The latter distribution in a non-magnetic insulator like PuB$_4$ would primarily reflect the extent of static bonding disorder. That we observed little change in the line shape between 90 and 690 days spectra for $B_0 \| c$ suggests that the amount of static disorder in the bonding network has not changed significantly within the first two years of aging or has little influence on the spectra after the first 90 days for this particular field orientation, reflecting an anisotropic hyperfine coupling between the orbital electrons and the $^{239}$Pu nuclei as discussed previously.[16]

To have a closer look into this point, we checked the field-orientation dependence of the single crystal spectrum after aging and compared it with the powder spectrum taken in a fresh sample.[16] Figures 2(a) and 2(b) present the field rotation patterns of the single crystal $^{239}$Pu-NMR spectrum at $T$ = 4 K for field rotation in the crystalline *bc*-plane [Fig. 2(a)] and *ab*-plane [Fig. 2(b)] taken after 640 days from the synthesis. Also shown at the bottom of Fig. 2(a) is the corresponding powder $^{239}$Pu-NMR spectrum of a polycrystalline sample from the same batch within the first week of its synthesis. There is an approximately single peak-like spectrum in the single crystal measurements for both rotation planes at all field orientations. The resonance line showed a sinusoidal angular dependence in the *bc*-plane rotation [see the inset of Fig. 2(a)], whereas little orientation dependence in the peak position was found in the *ab*-plane rotation. The overall orientation dependence suggests that the Pu ion sits in an axially symmetric environment along the crystalline *c*-axis and points to a dominantly axially symmetric Knight shift tensor.[17] This is consistent with our earlier consideration of local Pu-site symmetry being axial based on the line-shape analysis of the fresh-sample, powder-spectral pattern shown at the bottom of Fig. 2(a).

Notably, the maximum (at $B_0 \| b$) and minimum (at $B_0 \| c$) values of the Knight shift in the aged (640-days old) single-crystal nicely matches the corresponding values in the fresh (one-week old) powder sample [derived from the peak and kink positions indicated by the arrows in Fig. 2(a)]. In fact, by fitting the orientation dependence of the single-crystal spectral peak position in the *bc*-plane rotation with a form expected for axially symmetric systems,[32] $K = K_{\text{iso}} + K_{\text{ax}}(3\cos^2\theta - 1)$ [see the solid curve in the inset of Fig. 2(a)], we get $K_{\text{iso}}(4\text{ K}) = -0.099 \pm 0.008\%$ and $K_{\text{ax}}(4\text{ K}) = -0.488 \pm 0.005\%$, whose values are in excellent agreement with those extracted from the fresh polycrystalline powder spectrum.[16] [Here, $K_{\text{iso}}$ ($K_{\text{ax}}$) is the isotropic (axial) component of the Knight shift, and $\theta$ is the angle measured from the *c*-axis to the *b*-axis.] Furthermore, the linewidth in the single-crystal roughly accounts for the broadened tail structures seen at the two ends of the powder-NMR pattern.

These findings in Figs. 1 and 2 provide evidence that, from the perspective of the *static* NMR spectrum, aging effects that have developed and accumulated over 690 days have little impact on the axial site symmetry as well as on the magnitude and anisotropy of the $^{239}$Pu Knight-shift tensor. These features allow us to make a direct comparison of the spectral and shift data taken at different times from synthesis.

In Fig. 3(a) we show the temperature dependence of the Knight shift $K$ for $B_0 \| b$ in the aged single crystal (700-730 days old) plotted in comparison with the data for $B_0 \| c$ taken in the same sample but in a fresher condition (within 100 days, reported earlier in Ref. [16]). A monotonic decrease in the value of $K$ is observed for the two orientations towards lower temperature until $K$ levels off at around 80 K. The relative change from the low-temperature saturated value, $\Delta K$, shows activated behavior as highlighted in Fig. 3(b). To evaluate the excitation gap[15,16] in a most general way, we fitted $\Delta K$ with one of the commonly used forms in quantum spin systems,[18] $\Delta K = T^{-1}\exp\left(-\Delta_{\Delta K}/T\right)$. The deduced activation gap for the two orientations yield $\Delta_{\Delta K_c \cdot T} = 88.4 \pm 2.0$ meV for $B_0 \| c$ and $\Delta_{\Delta K_b \cdot T} = 89.4 \pm 19.1$ meV for $B_0 \| b$. The values of the two gaps are very close to each other, suggesting that the isotropic gap has not changed after two years. This provides support for the above arguments that there is qualitatively little impact of aging on the $^{239}$Pu NMR spectra at $T$ = 4 K, at least after the initial 90 days, which now is justified up to room temperature. (Other empirical fitting forms often employed in the analysis of $\Delta K$, such as $\Delta K = T^{-1/2}\exp\left(-\Delta_{\Delta K}/T\right)$ as summarized in Ref. [18], do not alter the extracted gap sizes and thus do not affect this argument.)

One interesting characteristic seen in the aged sample but omitted so far is that there is a slightly anisotropic feature in the spectrum that becomes obvious if we look more carefully into the spectral details for $B_0\|b$ and $B_0\|c$, as depicted in Fig. 4(a) [enlarged and replotted from Figs. 1(b) and 2(a)]. The symmetric shape of the spectrum for $B_0 \| c$ changes into a more complicated and asymmetric one for $B_0\|b$, with a couple of satellite-like peaks showing up at both sides of the main peak [labeled peaks S1 to S4 in Fig. 4(b)]. The presence of such structure inside of a spectrum indicates that there could be Pu nuclei sitting in slightly varying local environments. It is tempting to associate this additional line structure with radiation-damage induced alteration of the bonding network around Pu sites. However, as we don't have the corresponding data in a fresh sample, it is not possible to conclude whether this is due to aging or some other underlying structural defects and/or impurities that were present in the unaged sample. Either way, the effect seems to be rather small and more importantly the overall axial site symmetry is preserved at the $^{239}$Pu site. Therefore, to gain further insight into effects of aging, as a first approximation we omit this minor structural (or bonding) inhomogeneity and concentrate on the global nature of the spectrum, including NMR signal intensity and relaxation parameters ($T_1$ and $T_2$).

### B. Aging effects seen by $^{239}$Pu spin-lattice relaxation rate $1/T_1$ and spectral intensity

We turn to dynamic aspects of aging in the PuB$_4$ single crystal. Figure 5(a) presents the temperature dependence of the spin-lattice relaxation rate $1/T_1$ determined from the fully integrated intensity of the $^{239}$Pu-NMR spectrum for $B_0 \| b$ (recorded at 700-730 days from synthesis) and for $B_0 \| c$ (at 100 days, replotted from Ref. [16]). $1/T_1$ decreases rapidly upon cooling from room temperature for both field orientations akin to the Knight shift data in Fig. 3(a). An Arrhenius fit in the range $T$ > 150 K for $B_0 \| b$ gives an activation gap $\Delta_{1/T_1^b} = 80.0 \pm 7.2$ meV; whereas, a fit for $B_0 \| c$ and $T$ > 175 K yields an 80% larger value $\Delta_{1/T_1^c} = 143.7 \pm 9.1$ meV, as shown in Fig. 5(b). Below approximately 100 K, the temperature dependence of $1/T_1$ becomes qualitatively different between the two orientations. For $B_0 \| b$, $1/T_1$ tends to saturate as observed in the Knight shift $K$, but for $B_0 \| c$ it increases towards lower temperature and then decreases again below 20 K. Recalling that we do not see any notable anisotropy in the temperature dependence of $K$ (Fig. 3), the quantitative difference in gap values obtained from $1/T_1$ for the two orientations (Fig. 5) suggests that aging effects dominate as we are comparing results between aging after 700-days ($B_0 \| b$) and 100-days ($B_0 \| c$) of the same crystal. The different gap values extracted from $1/T_1$ and $K$ in the crystal after 100-days of aging gives a ratio of $\Delta_{1/T_1^c}/\Delta_{\Delta K_c \cdot T} \approx 1.63$, which is close to the average 1.73 for typical spin-gap systems.[18] In contrast, after the sample ages for 700 days, this same ratio determined with $B_0 \| b$ is about 0.9. (See Appendix B for further discussions.)

To better understand aging effects over an extended time period, we analyze all data taken on the PuB$_4$ single crystal over six years from its initial synthesis and make systematic comparisons, as summarized in Fig. 6. Specifically, we focus on three $^{239}$Pu-NMR quantities recorded at 5 K: (a) the full spectral signal intensity ($I$) times temperature ($T$), $I \times T$ [Fig. 6(a)]; (b) the spin-lattice relaxation time $T_1$ determined from stretched exponential fits to the nuclear relaxation curve [see Eq. (1); Fig. 6(b)]; and (c) the associated stretching exponent $\beta$ [Fig. 6(c)]. Here, we note that during this time period the sample experienced a complicated thermal cycling history: it was thermally cycled multiple times and also held at intermediate temperatures for various periods at temperatures between 4 K and 300 K, kept at 4 K for about a year within the first 500 days, and stored at room temperature in air on multiple occasions, in total more than four and half years (see Fig. S1). Despite this very complicated thermal history, we can clearly see in Fig. 6 a long-term underlying trend in all three NMR quantities as a function of time.

Figure 6(a) presents the first notable aging effect, which is a continuous overall loss of the integrated spectra intensity multiplied by temperature $I \times T$ as a function of time. (So-called $T_1$ and $T_2$ corrections are not needed here as our pulse conditions are so chosen that decays coming from $T_1$ and $T_2$ effects are negligibly small.) The value of $I \times T$ drops on average by 20% per year so that 40% of the signal is lost after two years from synthesis, and no signal at all remains in the fifth year and later. In NMR spectroscopy, the signal intensity $I$ is proportional to the size of the nuclear magnetization that follows the Boltzmann distribution of the nuclear Zeeman states and scales inversely to the temperature ($\propto 1/k_\text{B}T$). The intensity multiplied by $T$ compensates this Boltzmann factor, so $I \times T$ is thus proportional to the number of total nuclear spins probed by the measurements, which is usually a constant. The fact that $I \times T$ continuously decreases over time suggests that more and more nuclear spins become invisible in our measurement as aging proceeds. This loss of the $^{239}$Pu signal is not an experimental artifact because the reference $^{63}$Cu signal from the copper coil wrapping the sample shows no notable time-dependent change under NMR pulse conditions. We also note that $I \times T$ at various higher temperatures after 690 days from synthesis exhibits no $T$ dependence after making proper corrections (see Appendix A). This rules out any noteworthy intensity loss as a function of temperature (thermal wipeout) that often appears at low temperatures in various correlated electron systems that show magnetic ordering or glassy charge or spin freezing.[19-22]

A second aging effect manifests in the remaining signal and is characterized by a continuous change in the profile of the spin-lattice relaxation curve. Namely, the relaxation curve becomes increasingly stretched with increasing time after sample synthesis. Such a stretched feature in $T_1$ widely emerges in solids that possess certain types of electronic and/or structural inhomogeneity[18-21,23] and is known to represent a distribution in the value of $T_1$ across the sample that can be well characterized by a stretched exponential fit [Eq. (1)]. For instance, for $0.5 \leq \beta \leq 1$, the value of $1 - \beta$ represents the full width at half maximum of the distribution on a logarithmic scale, and the value of $T_1$ deduced from stretched fits stands for its median (note that we define $T_1$ as in Eq. (1) hereafter).[24] Figures 6(b) and 6(c) show the median $T_1$ and the exponent $\beta$ deduced from a stretched fit. Notably, $\beta$ [Fig. 6(c)] continuously decreases with time and reaches $\beta \approx 0.5$ at 730 days, which corresponds to an order of magnitude distribution of $T_1$.[24] The median value of $T_1$ [Fig. 6(b)] seems to be more thermal history dependent than $\beta$ and $I \times T$. Nevertheless, a systematic aging test performed at 5 K, as summarized in Fig. 7, captures an increasing trend of $T_1$ with aging at low temperature (see Sec. III.D for details).

Before moving into the details of the changing dynamics of PuB$_4$ with aging, let us first reiterate that there are two dynamical manifestations of the aging effect that is found here. The first one is the overall loss of the $^{239}$Pu-NMR signal intensity upon aging, and the other one is the developing inhomogeneity in the remaining signal that shows up as a growing distribution in the $T_1$ relaxation time across the sample when it is held at base temperature.

The intensity loss typically might be associated, at least in part, to disorder induced by alpha decay of $^{239}$Pu atoms that produces a $^{235}$U recoil atom as well as a helium nucleus ($\alpha$ particle). The dominant source of radiation damage is known to come from the heavier uranium recoil atoms.[1] In elemental $\delta$-Pu the recoil uranium atoms trigger a concentrated displacement damage around stopping sites by knocking off the nearby Pu atoms from their original lattice positions. Wolfer[1] estimated the accumulation rate of such displacement damage in $\delta$-Pu to be 0.068 displacement per $^{239}$Pu atom per year, which means each Pu atom is displaced from its equilibrium position on average in 14.7 years. This displacement rate scales inversely with the sample's melting temperature,[1] which for $\delta$-Pu is ~ 500 K and for PuB$_4$ is ~ 2,400 K,[25] so we expect approximately 0.01 displacement per $^{239}$Pu atom per year in PuB$_4$. If we use this estimated displacement rate in PuB$_4$ and assume that displacements suppress the Pu signal by inducing changes in the static and dynamic local hyperfine environment (leading to an excess shift and too-fast relaxation at the displaced Pu atoms), we would expect that, without annealing the damage, only about 2 % of the signal should be lost after two years of accumulated site displacements. Instead, 40 % of the signal is lost in this period, as seen in Fig. 6(a), which includes excursions to room temperature (Fig. S1) where some damage should be "healed". Though other self-damage routes must contribute to a loss of signal, it is extremely unlikely that they could account for a more than order-of-magnitude discrepancy between estimated and observed signal loss. The unexpected excess loss suggests that the consequence of radiation damage is not just simple static displacements of the atoms but has a more spatially extended dynamic influence. A growing distribution in $T_1$ in the spectrum together with this excess signal-intensity loss is consistent with a fluctuating hyperfine field developing and distributing non-uniformly across the sample being a contributing factor. We will discuss such dynamic aspects of aging in more detail in the following Secs. III.D and IV.

## C. Isochronal annealing effect on $^{239}$Pu intensity and $T_1$

In addition to the spectral and relaxation measurements discussed so far, we have also examined the outcome of isochronal annealing of the PuB$_4$ single crystal within the first two years after its synthesis.

The annealing test was conducted three times: The sample underwent its first annealing at day 370 after being kept at 5 K over approximately 140 days during which it accumulated radiation damage at low temperature. The second annealing was done at day 400 after keeping the sample at 5 K for another month. Following these two annealing tests, the sample was stored at room temperature in air for over 140 days and was used for measurements at varying temperature conditions between 4 K and 250 K to collect temperature-dependence data. The third annealing test was then conducted at day 690.

For the first two trials, we annealed the sample at $T_A$ = 300 K for three hours, then cooled back to 5 K and performed NMR measurements with $B_0 \| c$. For the third trial, we did a systematic incremental annealing in a way such that we first stayed at 4.5 K, raised the sample temperature to a target value of $T_A$ (= 30, 60, 100, 150, and 250 K) and stayed there for an hour, came back to 4.5 K and did NMR measurements with $B_0 \| b$, and then moved to the next higher target $T_A$ and repeated the sequence of hold, cool, and measure.

Figure 7 summarizes the first and second isochronal annealing test results, in which we show the full-spectral NMR signal intensity $I \times T$ [Fig. 7(a)], the median spin-lattice relaxation time $T_1$ [Fig. 7(b)], and the stretching exponent $\beta$ [Fig. 7(c)] as a function of time. Arrows in the figure indicate the two annealing tests. Before moving into detailed influences of annealing, we reiterate that cryogenic damage accumulations are observed in these quantities. Over the time at 5 K (day 240 to 380 and day 400 to 440), we see a systematic decrease in $I \times T$ and $\beta$ (within error bars) and an increase in the median value of $T_1$ as damage builds up. For the first annealing test all three quantities show a remarkable change after annealing that tends to reduce the accumulated damage. In fact, $I \times T$ recovers by 10 %, the median $T_1$ shortens by 30 %, and $\beta$ increases from 0.63 to 0.68, suggesting that the effects of aging have been partially healed by isochronal annealing. The second attempt was, however, not effective, and we see only a minor influence of annealing on the NMR observables. In Fig. 8 we summarize results of the third annealing test conducted a year later. In both the spectrum [Fig. 8(a)] and the spin-lattice relaxation curve [Fig. 8(b)], no change was observed after annealing, suggesting that the accumulated effects of aging are becoming more difficult to remove. That we found a clear impact of annealing in the first year but no effective annealing in the second year implies that there may be a threshold of accumulated damage beyond which annealing at a $T_A$ of 250-300 K is only partially effective in healing damage.

$PuB_4$ is a refractory material, with strong bonding characterized by a high melting point of nearly 2,430 K.[25] It is likely that much higher temperatures are needed to thermally heal the sample with isochronal annealing. Unfortunately, for safety reasons we were unable to reach such temperatures. Materials with similar bonding, such as actinide oxides, have demonstrated beneficial annealing at ~ 1,500 K, which restored lattice parameters to their undamaged values.[26] In contrast, actinides with metallic bonding, such as Ga stabilized $\delta$-Pu metal, show responses to annealing at relatively low temperatures.[2] Similar annealing effects at temperatures below 300 K have also been observed in irradiated metals.[27] Together, this suggests that the nature of bonding in a material determines the temperature needed to "heal" the radiation-induced damage. As we have demonstrated, actinide NMR is sensitive to isochronal annealing in a strongly bonding material like $PuB_4$ in its early stages of damage accumulation, and this study encourages more controlled aging studies of plutonium compounds using direct [239]Pu NMR.

**D.** *Microscopic inhomogeneity and fluctuating hyperfine fields probed by $^{239}$Pu $1/T_1$ and $1/T_2$*

Additional insight into the effects of aging comes the variation of $1/T_1$ and $1/T_2$ across the spectrum after 640 days from synthesis. For this, we focus on these relaxation rates that are derived from a fraction of the spectral intensity obtained by integrating a small slice of spectrum (a window size of 0.02 % of $K$). At that time, the full spectrum already possesses approximately an order of magnitude distribution in the size of $1/T_1$ [corresponding to the stretching exponent of $\beta = 0.5$, see Fig. 6(c)], and the full spectral intensity has dropped by 40 % from that in a fresh sample [Fig. 6(a)].

Figure 9 shows the distribution of $1/T_1$ and $1/T_2$ across the spectrum for $B_0 \| b$ at $T$ = 4 K. Interestingly, the value of (median) $1/T_1$ does not change appreciably across the spectrum, yet at each point it showed very stretched behavior with the stretching exponent $\beta$ being close to 0.5. This stretched feature is exactly the same as we deduced from analysis of the full spectrum and indicates a distribution of $T_1$ that reflects a microscopically inhomogeneous distribution of fluctuating hyperfine fields. This heterogeneity arises irrespective of a spatial distribution of static hyperfine fields (i.e., the underlying bonding network) that determines the single-crystal spectral shape.

The dephasing nature of the spin-echo signal gives further insight into this issue. $1/T_2$ in Fig. 9 shows a sizable variation across the spectrum and compared to $1/T_1$ presents a very different evolution as a function of $K$. The functional form of the spin-echo decay curve is purely exponential [i.e., $\alpha \approx 1$ in Eq. (2)] at the low-shift end of the spectrum (at $K = 0.24$ %), whereas it becomes more Gaussian-like (although not quite purely Gaussian) across the main part of the spectrum with $\alpha \approx 1.4 - 1.8$. The spin-spin relaxation process, which determines $1/T_2$, is generally related to dephasing of the spin-echo signal due to a magnetic dipole interaction between nuclear spins and the dynamic hyperfine interaction between electrons and nuclear spins.[28] The size of the former contribution can be estimated via van Vleck's second moment technique[17] and leads, in PuB$_4$, to temperature-independent values of $1/T_2 = 4.4 \times 10^{-6}$ msec$^{-1}$ for like spins ($^{239}$Pu–$^{239}$Pu) and $1.8 \times 10^{-2}$ msec$^{-1}$ for unlike spins ($^{239}$Pu–$^{11}$B). These values are remarkably smaller than the experimental $1/T_2$ values in Fig. 9, indicating that the dominant contribution to spin-echo dephasing comes from the latter dynamic hyperfine interaction. As the variation profile of $1/T_2$ across the spectrum differs significantly from that of $1/T_1$, we speculate that there is a distribution in the spectral density of fluctuating hyperfine fields that brings contrasting impacts on $1/T_1$ and $1/T_2$ relaxation mechanisms.

The temperature dependence of $1/T_2$ should offer further insights about inhomogeneity, but when we attempted to measure spin-echo decay in the period 690-700 days after synthesis, the intensity loss due to aging prevented standard $T_2$ measurements, except at low temperatures (Fig. S2). Instead, we estimate $T_2$ at various temperatures in a non-standard way. Here we use the temperature dependence of the bare $I \times T$ that was determined from full spectral intensities with a fixed waiting time $\tau$ = 100 μs between $t_{90°}$ and $t_{180°}$ pulses (Fig. S3). The estimate relies on our observation that the $T_2$-corrected $I \times T$ at 4.2 K has the same magnitude as at 100 K. (Specifically, the $T_2$ correction allows us to compensate for $T_2$-decaying effects [Eq. (2)] and recover the intensity for an imaginary infinite-$T_2$ limit). The bare $I \times T$ shows strong temperature dependence below 60 K and at 4.2 K becomes half of its value at high temperatures (Fig. S3). This suggests that the apparent decrease of the bare $I \times T$ arises from a shorting of $T_2$. Assuming a temperature-independent stretching exponent in Eq. (2) and its value at 4.2 K ($\alpha$ = 1.52), we use Eq. (2) to convert the observed intensity in Fig. S3 to corresponding estimates of $T_2$ at different temperatures, which are dubbed $T_{2,\text{eff}}$ hereafter (see Appendix A for details). The value of $T_{2,\text{eff}}$ is strictly speaking not identical to the true $T_2$ but seems to provide its reasonable representation.

Figure 10(a) presents the thus determined temperature dependence of $1/T_{2,\text{eff}}$ with $B_0 \| b$. At high temperatures, $1/T_{2,\text{eff}}$ is weakly temperature dependent as expected for dephasing of nuclear spins by magnetic dipole interactions. With decreasing temperature, however, $1/T_{2,\text{eff}}$ increases notably below 40 K and at the base temperature reaches five times its high-temperature value. This strong temperature dependence of $1/T_{2,\text{eff}}$ points to the underlying trend of the true $1/T_2$ that equally grows toward low temperatures. Note that in the same temperature range, the linewidth of the spectrum does not change significantly as a function temperature [Fig. 10(b)]. The characteristic increase of $1/T_{2,\text{eff}}$ cannot be driven by the nuclear dipole interaction and combined with little change in the linewidth strongly suggests that fluctuations of the electronic hyperfine field should be responsible for shortening $T_2$. Interestingly, there is no increase of $1/T_1$ in the corresponding low temperature range [inset of Fig. 10(a)]. This observation implies that there should be a fluctuating hyperfine field that does not affect the $T_1$ relaxation process but strongly influences the $T_2$ relaxation process in the aged PuB$_4$ sample.

## IV. DISCUSSION

The single crystal [239]Pu-NMR reveals dynamic spin fluctuations that cause a notable distribution of fluctuating hyperfine fields across the sample and that they could be equally responsible for the signal wipeout with little change in the line shape that develops over an extended time period. The shortening of $T_2$ seems to be critical for all of these observations and reflects a time-scale of the longitudinal component (parallel to the external field $B_0$) of the spin fluctuations near zero frequency.

The NMR spin-spin relaxation rate $1/T_2$ can be generally decomposed into nuclear and electronic contributions: $1/T_2 = (1/T_2)_\text{n} + (1/T_2)_\text{el}$. The latter dominates the $T_2$ relaxation process in PuB$_4$ [i.e., $(1/T_2)_\text{n} \ll (1/T_2)_\text{el}$] as discussed in Sec. III.D. The electronic part can be further written as a sum of two distinct contributions in terms of the auto-correlation function of the fluctuating hyperfine field $\mathbf{h}(t)$ as[17]

$$(1/T_2)_{\text{el}} = (1/T_2)_{\text{el}}^\perp + (1/T_2)_{\text{el}}^\parallel, \qquad (3)$$

$$(1/T_2)_{\text{el}}^\perp = \gamma^2 \int_{-\infty}^{\infty} dt \, \langle h_\perp(t) h_\perp(0) \rangle \exp(i\omega_L t) = A_\perp(\omega_L),$$

$$(1/T_2)_{\text{el}}^\parallel = A_\parallel(0). \qquad (4)$$

The spin-spin relaxation rate thus picks up fluctuations perpendicular to $B_0$ at the NMR Larmor frequency ($\omega_L$) [represented by $A_\perp(\omega_L)$] and is also sensitive to the longitudinal fluctuations parallel to $B_0$ near zero frequency [given by $A_\parallel(0)$].[28] The first term is known to be driven by spin-lattice $T_1$ relaxation processes and leads to the so-called Redfield contribution to $1/T_2$, which for a spin $I$ = 1/2 system is given by $(1/T_2)_{\text{el}}^\perp = (1/T_1)/2$.[28-30] With $1/T_{2,\text{eff}} \gg (1/T_2)_{\text{el}}^\perp$ over the entire temperature range as seen in the inset of Fig. 10(a), our data naturally suggest the growing $1/T_{2,\text{eff}}$ (and $1/T_2$) towards low temperature to be dominantly caused by longitudinal fluctuations near zero frequency [$(1/T_2)_{\text{el}}^\parallel$].

We recall that the so far discussed $T_{2,\text{eff}}$ and $T_2$ is averaged over the full spectrum, and in aged PuB$_4$, $T_2$ strongly distributes across the spectrum at least at base temperature, as seen in Fig. 9. When only a fraction of the spectrum is excited on the lower frequency side where $1/T_2$ is largest, the functionality of the echo-decay curve becomes purely exponential. Such an exponential decay is characteristic of fluctuations of the longitudinal component of the hyperfine field $h_\parallel$[31] in the fast motion regime, $\gamma \langle h_\parallel^2 \rangle^{1/2} \tau_c \ll 1$, and the corresponding spin-spin-relaxation rate is given by[30]

$$(1/T_2)_{\text{el}}^\parallel = \gamma^2 \langle h_\parallel^2 \rangle \tau_c, \qquad (5)$$

where $\tau_c$ is the correlation time of the fluctuating hyperfine field that is defined by an exponentially decaying auto-correlation function as $\langle h_\parallel(t) h_\parallel(0) \rangle = \langle h_\parallel^2 \rangle \exp(-t/\tau_c)$. The increasing trend of $1/T_{2,\text{eff}}$ in Fig. 10(a) thus indicates a rapid increase of $\tau_c$ with decreasing temperature. Importantly, as $T_2$ is strongly distributed across the sample, the fluctuations should constitute a finite spectrum with a certain width. The spectral intensity loss we observe, therefore, could be due to the low frequency tale of this spectrum in which $\tau_c$ gets so large that the time scale of $T_2$ becomes shorter than the dead time of the measurement, resulting in the signal wipeout.

Growing slow fluctuations of such a longitudinal hyperfine field have been reported in a variety of cases. For example, similar fluctuations were observed in classical and quantum spin[32-34] and charge[35,36] glasses, ionic and molecular motion,[31,37,38] charge- and spin-ordered cuprates,[19,39,40] frustrated[41,42] and quantum critical[43] magnets, as well as disordered Mott insulators[44,45] and nematic domain walls.[46,47] In these cases, spatial inhomogeneity seems to be required for the fluctuations to develop.

For a Pu-based material like PuB$_4$, self-irradiation damage due to radioactive decay of Pu atoms[1] is the most plausible source of inhomogeneity, which can cause accumulated magnetic and non-magnetic defects upon aging.[48] In our case, $\alpha$-decay of $^{239}$Pu (93.96 weight %, half-life 24,110 years) should be one of the main causes. This decay channel generates thousands of non-magnetic lattice defects (Frankel pairs) per decay[1] and $^{235}$U recoil atoms as a daughter product at a rate of $\approx$25 ppm/yrs. The $\beta$-decay of $^{241}$Pu (0.025 weight %, half-life 14.329 years) is another source that results in an accumulation of $^{241}$Am atoms at an even faster rate of $\approx$40 ppm/yrs. Both of these recoil atoms can be radicals that behave as paramagnetic centers[48] that from ligand NMR are known to cause a shortening of $T_2$ and in turn a dynamic NMR signal wipeout[49] at nearby ($^{239}$Pu) nuclei due to the direct dipole-type hyperfine coupling that provides fluctuating hyperfine field at the target nuclei. However, the amount of radicals and the number of nearby [239]Pu atoms feeling the fluctuating field seems to be way too small to account for the observed massive intensity loss in [239]Pu NMR that develops as fast as $\approx -20$ %/yrs. This notable discrepancy suggests that there are some indirect hyperfine coupling processes that boost the signal wipeout.

For a metallic system, one possible origin for this discrepancy could be the indirect relaxation between radical spins and host nuclear spins that is assisted by conduction electrons. This type of indirect hyperfine interaction is a nuclear spin version of the Ruderman-Kittel-Kasuya-Yosida (RKKY) interaction. It drives an oscillatory interaction that makes the influence of radical-induced electron spin polarization spatially more extended (see for example Ref. [50]). In magnetic dilute alloys the impact of such an indirect interaction on the NMR relaxation parameters and intensity has been extensively studied and is shown to cause a sizeable shortening of $T_1$ and $T_2$ even for a tiny magnetic impurity concentration of a few hundreds of ppm or less per unit cell.[50-55] The observed pronounced intensity loss could be, therefore, due to this sort of indirect (RKKY) interaction that enhances the wipeout effect over time.

Questions remain though. Perhaps the most immediate question is whether there are electrons within the gapped energy spectrum of PuB$_4$ that would allow the indirect RKKY-type hyperfine interaction. Although there is no concrete answer to this question, our conjecture is supported by the proposition of in-gap states in PuB$_4$ suggested earlier by transport and heat capacity[15] as well as NMR $1/T_1$[16] measurements at temperatures below the energy scale of the gap (below 100 K). Certainly, in-gap states have been widely observed in a group of related narrow-gap Kondo insulators, such as SmB$_6$ and YbB$_{12}$.[56-58] In these, impurities and defects play a role in producing such low-energy states as well as in their bulk and surface electronic properties.[59,60] We speculate that radioactive decay in PuB$_4$ not only generates magnetic radicals but also metallic impurity states that mediate the indirect hyperfine interaction. The effects of aging in this context could be understood to appear as a shift of the fluctuation spectrum towards lower frequency that turns more and more of the sample into the wipeout regime. Similar aging effects have been previously observed in AmO$_2$ by ligand $^{17}$O NMR.[8] Further systematic investigations with X-ray diffraction and radiochemical analysis would be useful to improve our understanding of the complex mechanism of radiation damage in this Pu-based material.

## V. CONCLUSIONS

We have used $^{239}$Pu-NMR to spectroscopically characterize consequences of aging on static and dynamic properties of single-crystal PuB$_4$. Field-angle dependent measurements imply an effective tetragonal site symmetry of Pu atoms despite their sitting in an orthorhombic crystalline environment. The $^{239}$Pu resonance spectrum of aged PuB$_4$ reveals the existence of multiple, slightly distinct Pu sites possessing very similar axial local-site symmetry. Aging has little effect on the static NMR spectrum, whereas it shows up more directly and strongly in the spectral intensity as well as in the dynamic spin-lattice relaxation ($T_1$) and spin-spin relaxation ($T_2$) profiles. Aging drives $T_1$ to longer values, increases its distribution across the sample, and reduces the spectral intensity by 20 % per year. There is a strong variation in $T_2$ over the spectrum and a notable shorting of $T_2$ below about 40 K, albeit with no appreciable change in $T_1$. Together, these point to the development of slow fluctuations that could be responsible for wipeout of the spectral signal over the course of years. Isochronal annealing at room temperature only partially heals the aging damage in the first year and has little effect after a second year of accumulated damage, suggesting the presence of a threshold amount of damage that is healable. This work demonstrates that the actinide resonance spectroscopy can be used as a uniquely sensitive probe to study radiation damage in materials.

## ACKNOWLEDGEMENTS

We thank H. Yasuoka, H. Mason, A. Altenhof, and J. Thompson for fruitful discussions and comments on aging effects and relaxation rate analyses as well as on the manuscript. The dynamic nuclear spin-lattice relaxation measurements were performed under the auspices of the U.S. Department of Energy, Office of Basic Energy Sciences, Division of Materials Science and Engineering. The spectral measurements including their rotation and aging studies were supported from the Laboratory Directed Research and Development programs and LANL Office of Experimental Sciences. We additionally thank the U.S. Department of Energy, Office of Science, Office of Basic Energy Sciences, Heavy Element Chemistry program (2020LANLE372) for funding (SAK).

### *Appendix A. Determination of temperature dependence of $1/T_{2,\,eff}$ from intensity data*

Figure S1 shows the $^{239}$Pu-NMR spin-echo decay curve recorded for $B_o \parallel b$ at base temperature (4.2 K) in a sample after 690 days from its synthesis. In Fig. S3 we present the temperature dependence of the full-spectral NMR signal intensity $I \times T$ recorded by using a fixed waiting time $\tau$ (= 100 μs) between $t_{90°}$ and $t_{180°}$ pulses in the spin echo with a last delay at least ten times longer than $T_1$. A nearly temperature independent $I \times T$ is seen at higher temperature, which is expected if the number of the nuclear spins excited by the NMR pulses is preserved. With decreasing temperature, $I \times T$ shows a strong decrease and at base temperature is only about 50 % of its higher temperature value. This clear intensity drop is not a signal wipeout: after a proper $T_2$ correction, based on the spin-echo decay curve in Fig. S2, the value of $I \times T$ perfectly matches the high temperature constant value. As PuB$_4$ is a modestly good insulator and will not be accompanied by a skin-depth effect, this intensity decrease should represent a shortening of $T_2$. Although we don't have enough $T_2$ data at various temperatures, this intensity data allows us to estimate the value of $T_2$, dubbed $T_{2,\,\text{eff}}$ hereafter, at different temperatures by assuming that the functional form of the spin-echo decay curve [Eq. (2) in the main text] does not differ from that at 4.2 K (Fig. S2):

$$\frac{1}{T_{2,\,\text{eff}}} = -\frac{1}{2\tau}\left(ln\left[\frac{I \times T}{(I \times T)_0}\right]\right)^{1/\alpha} \tag{S1}$$

where $I \times T$ is the intensity at a given temperature, $(I \times T)_0$ is the $T_2$-corrected constant intensity at 4.2 K, $\tau = 100$ μs, and $\alpha = 1.52$ is the exponent determined at 4.2 K.

### Appendix B. Fitting forms to the temperature dependence of $\Delta K$ and $1/T_1$

Ref. [16] reported identical results for $\Delta K$ [$B_0 \| c$ data in Fig. 3(a)] and $1/T_1$ [$B_0 \| c$ data in Fig. 5(a)] using other fitting functions and obtained bigger gap sizes of $\Delta_K$ = 155.6 meV and $\Delta_{T_1}$ = 251.3 meV. That work assumed a model density of states, following Caldwell et al[57] for their analysis of the $^{11}$B-NMR results in SmB$_6$, and employed the so-called Korringa relation[17,30] that connects the NMR Knight shift and $1/T_1$ via electronic density of states at the Fermi level $E_F$. This analysis explicitly assumes that the deduced gap in the NMR measurements reflects the size of the band gap around $E_F$ (as NMR in this case probes the number thermally excited electron-hole pairs generated across the band gap). Moreover, it relies on multiple fitting parameters that control the shape of the model density of states used in that fit. Although that analysis allowed the authors to make a direct quantitative comparison of the gap sizes with those reported in SmB$_6$, it is not obvious whether a simple semiconductor-like picture is relevant to PuB$_4$, where the itinerant versus localized picture of 5f-electrons is not well understood. To avoid unwanted complications, we instead have taken a more general approach and used commonly employed exponential forms for spin-gapped systems[18] in the analysis of the spin gaps in Figs. 3 and 5. Despite the different approaches used to extract the gaps, the ratio of the two gap, $\Delta_{T_1}/\Delta_K$, is almost identical, within estimated error, to that reported in Ref. [16]. This common ratio supports the validity of our approach.

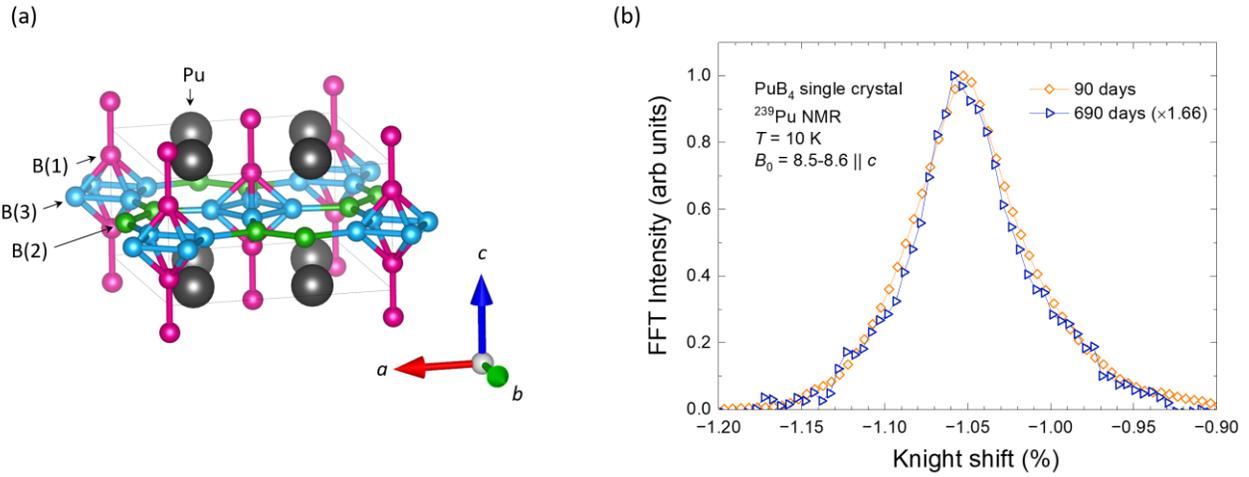

**FIG. 1.** (a) Tetragonal crystal structure of PuB$_4$. Grey balls are plutonium atoms (having locally orthorhombic symmetry), and pink [B(1)], green [B(2)], and blue [B(3)] balls are the three crystalographically nonequivalent boron atoms in the unit cell (given by the light grey box). The colored lines between two atoms are the bondings connecting the nearest neighbor atoms. (b) Typical $^{239}$Pu NMR spectrum in a single crystal of PuB$_4$ under an external magnetic field of $B_0$ = 8.5-8.6 T applied parallel to the crystalline $c$-axis ($B_0 \| c$). Diamonds are the data taken in a fresh sample (90 days from synthesis), whereas the triangles give the corresponding data taken in the identical sample 600 days later (690 days from synthesis).

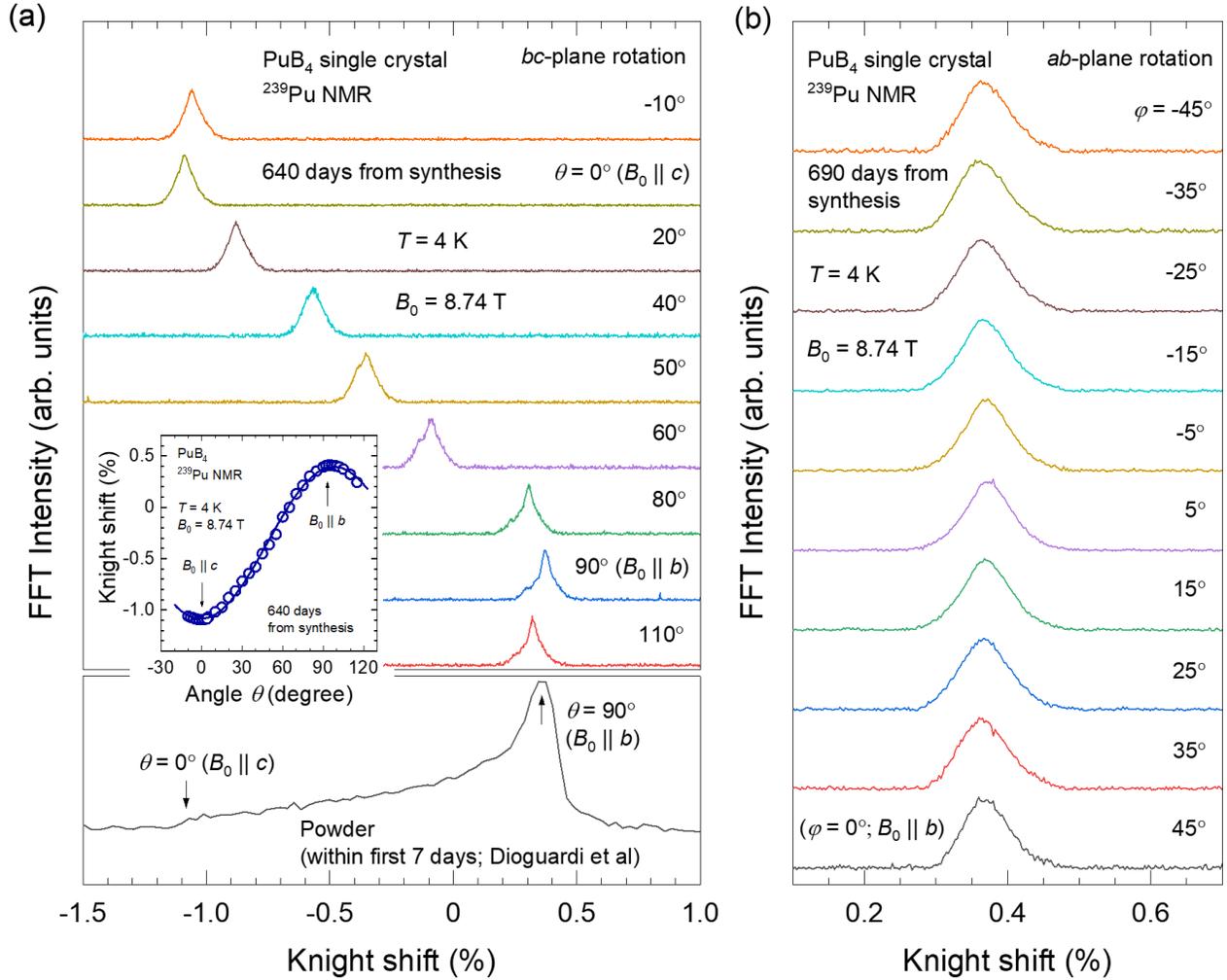

**FIG. 2.** Field orientation dependence of $^{239}$Pu-NMR spectrum in PuB$_4$ single crystal at 4 K. The magnetic field $B_0$ of 8.74 T is rotated in the crystalline $bc$-plane in (a) and in the $ab$-plane in (b). The angle $\theta$ ($\varphi$) is defined as the direction of $B_0$ measured from the crystalline $c$-axis to $b$-axis ($b$-axis to $a$-axis). The data are taken after 640 days from synthesis in (a), and after 690 days in (b). Note that the small peak-like structures seen for $\theta \approx 90°$ around the sharp central peak in (a) [labeled S1-S4 in Fig. 4(b)] are absent in (b) because we used 33% smaller excitation window in the pulsed NMR measurements to predominantly excite the sharp peak and focus on the angular ($\varphi$) dependence of its center. The bottom spectrum in (a) is the corresponding powder spectrum in a fresh polycrystalline sample from the same batch taken within the first week from synthesis (replotted from Ref. [16]). Inset in (a) shows the field orientation dependence of the Knight shift value at the sharp peak position of the spectrum in the $bc$-plane plotted as a function of the angle $\theta$. Solid curve gives a sinusoidal fit to the data points (see the text).

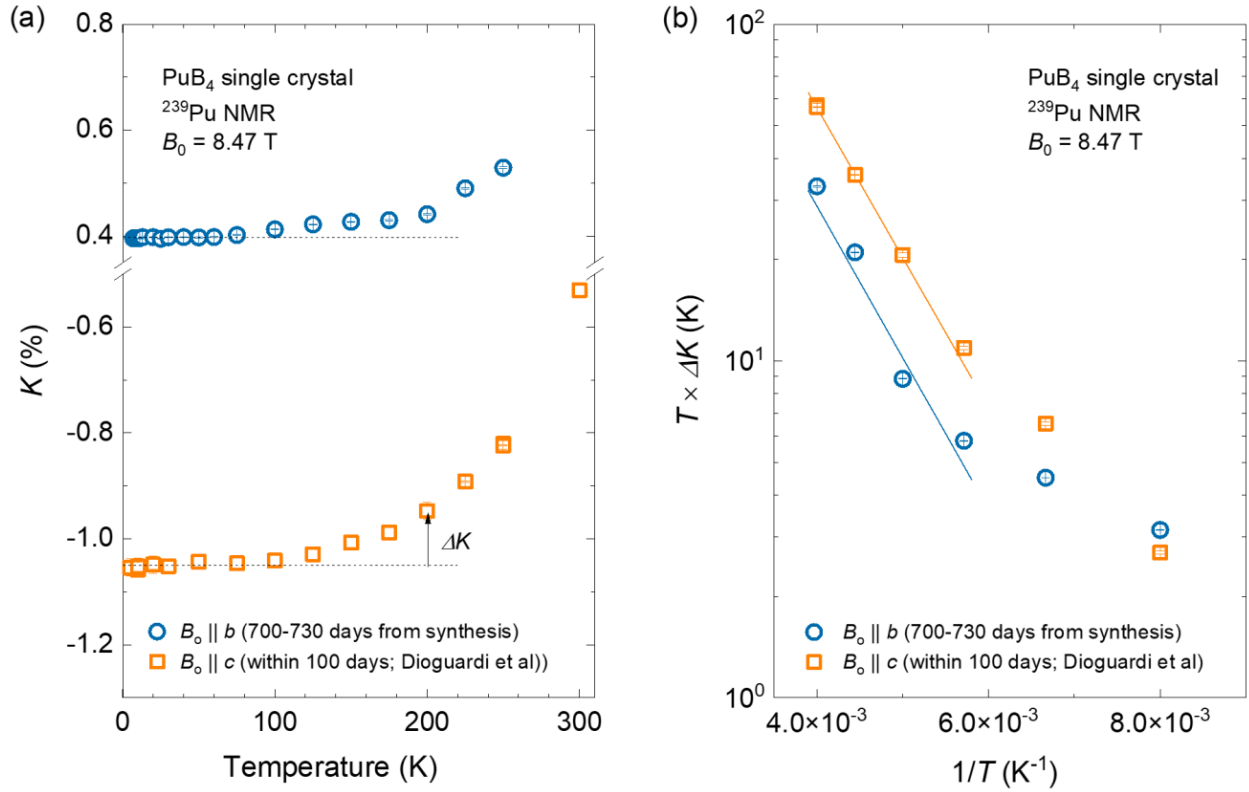

**FIG. 3.** (a) Temperature dependences of $^{239}$Pu Knight shift $K$ in PuB$_4$ single crystal at $B_0$ = 8.47 T. The results for $B_0 \| c$ (circles) are taken at 100 days from synthesis (replotted from Ref. [16]), and those for $B_0 \| b$ (squares) are recorded at 700-730 days. Dashed lines represent the leveled off values of the shift at low temperature below 80 K, and $\Delta K$ is defined as the high temperature values measured from that saturated value for each orientation. (b) Arrhenius plot of $\Delta K \times T$ plotted against inverse temperature. Solid lines are exponential fits to the data above 200 K with the form $\Delta K = T^{-1} \exp(-\Delta_{\Delta K}/T)$ (see the text).

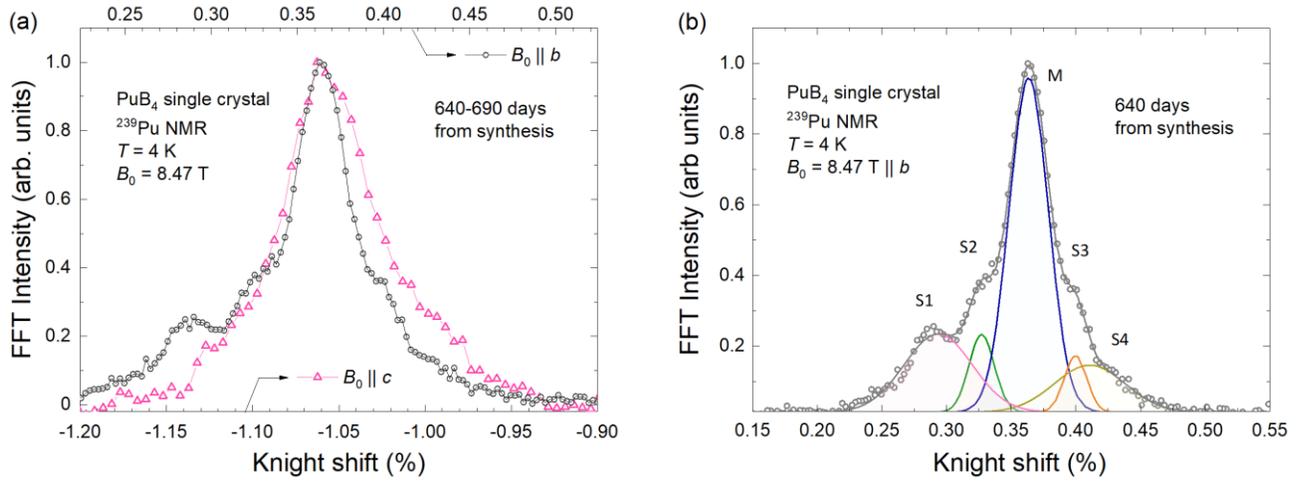

**FIG. 4.** (a) Comparisons of the $^{239}$Pu-NMR spectrum for $B_0\|b$ and $B_0\|c$ taken after 640-690 days from synthesis [replotted from Fig 2(a)]. The two curves are shifted horizontally to show them on top of each other. (b) Spectrum for $B_0\|b$. In addition to the central sharp peak (labeled M), a couple of peak-like features can be resolved in both sides of the spectrum (labeled S1 to S4). Solid curves stand for five Gaussian fits to the spectrum.

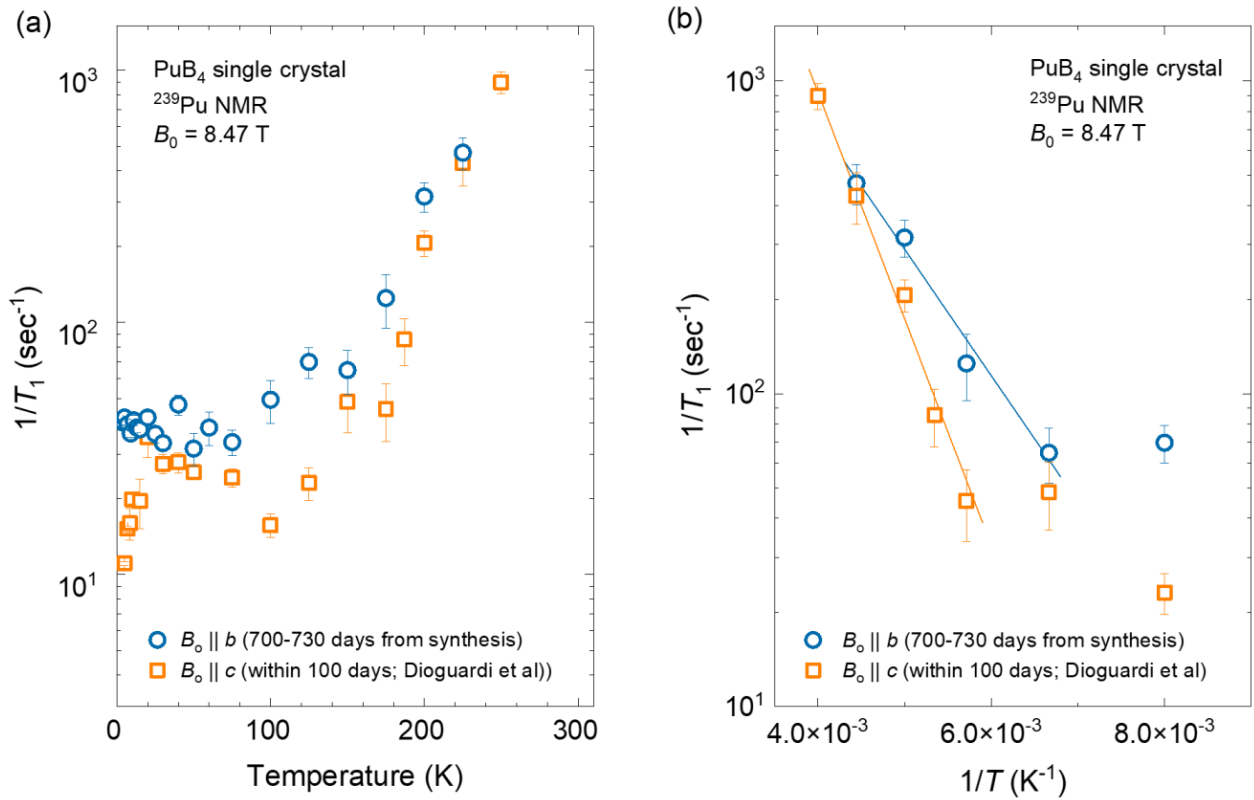

**FIG. 5.** (a) Temperature dependences of $^{239}$Pu nuclear spin-lattice relaxation rate $1/T_1$ in PuB$_4$ single crystal at $B_0 = 8.47$ T. The results for $B_0 \| c$ (circles) are taken at 100 days from synthesis (replotted from Ref. [16]), and those for $B_0 \| b$ (squares) are recorded at 700-730 days. (b) Arrhenius plot of $1/T_1$ versus inverse temperature. Solid lines are exponential fits to the data above 200 K (see the text).

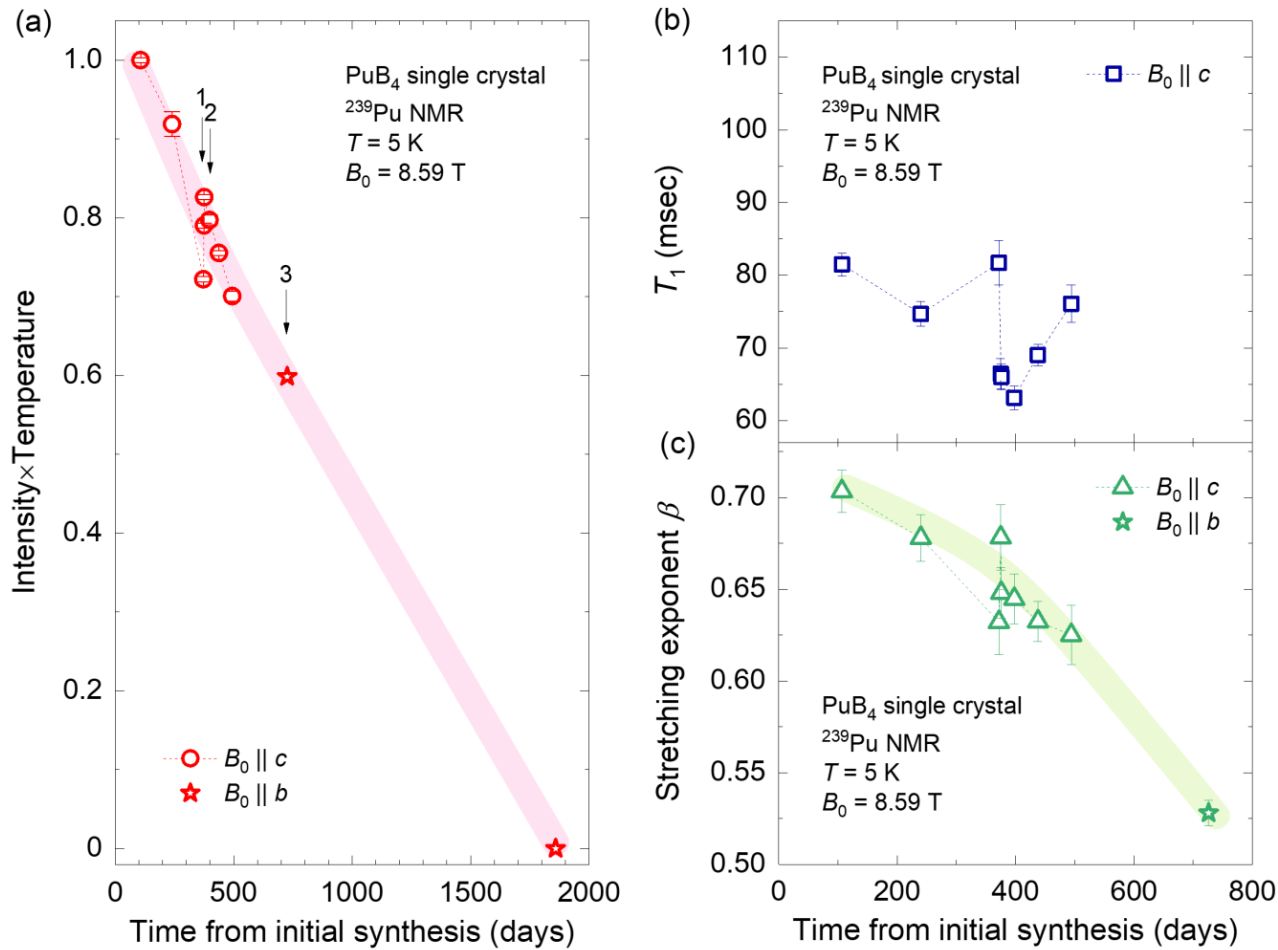

**FIG. 6.** Direct $^{239}$Pu-NMR representation of aging effect in PuB$_4$ single crystal within three years from synthesis. (a) Integrated spectral intensity times temperature. Arrows indicate three isochronal annealing tests performed at 370 and 400 days (first and second trials, respectively, in Fig. 7) and at 690 days (third trial in Fig. 8) from initial synthesis. No signal was detected after 1850 days. (b), (c) Nuclear spin-lattice relaxation time $T_1$ of the integrated intensity (b) and the corresponding stretching exponent $\beta$ of the nuclear relaxation curve (c). The $T_1$ data point at 720 days is not shown in (b) as it has different anisotropic factor than all the other points and is not directly comparable. That influence seems to be relatively small for $\beta$, and thus the corresponding point is present in (c). Dashed and shaded lines are guides to the eyes.

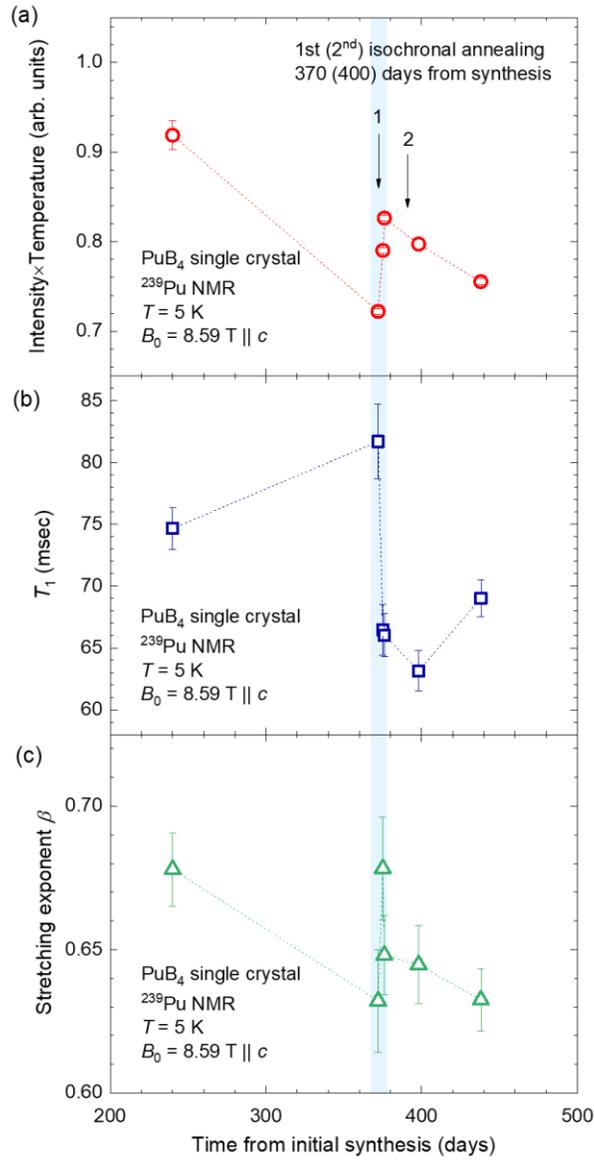

**FIG. 7.** First and second isochronal annealing tests probed by $^{239}$Pu NMR in PuB$_4$ single crystal within the first year from synthesis. Plotted data are (a) the integrated spectral intensity times temperature, (b) the median nuclear spin-lattice relaxation time $T_1$, and (c) the stretching exponent $\beta$ of the corresponding $T_1$ nuclear relaxation curve. Between annealing tests, the sample was kept at base temperature (5 K) from day 240 to 380 and from day 400 to 440 at $B_0$ = 8.59 T for $B_0 \| c$. The first annealing was then conducted at day 370 [indicated by an arrow with the label 1 in (a)], at which we raised the sample temperature to 300 K, stayed there for three hours, came back to the base temperature, and performed NMR to check the impact of annealing. We then kept the sample at 5 K for one more month and then conducted the second annealing at day 400 [indicated by the second arrow in (a)].

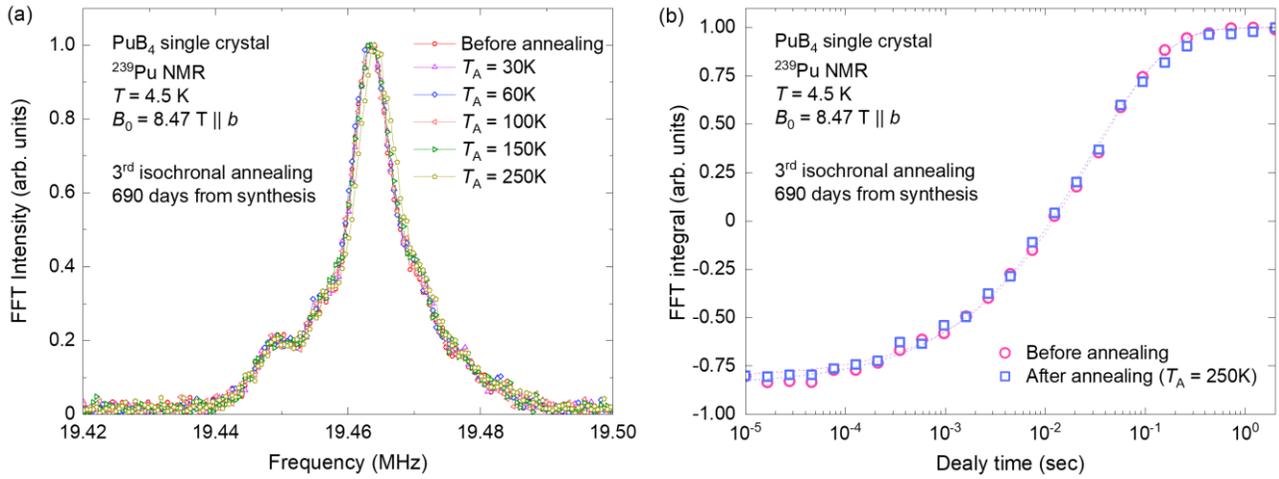

**FIG. 8.** Third isochronal annealing test after 690 days from synthesis. The sample was stored at room temperature after the first two annealing tests in Fig. 7 from day 500 to day 640 and then kept at various low temperature conditions from 4 K to 250 K until performing the third annealing test at day 690 [see also Fig. 6(a)]. (a) $^{239}$Pu-NMR spectrum at 4.5 K at $B_0$ = 8.47 T for $B_0 \| b$ after isochronal annealing performed at various temperatures. Different symbols represent the spectrum recorded after keeping the sample at the annealing temperature $T_A$ for an hour and coming back to 4.5 K. (b) Comparison of nuclear spin-lattice relaxation ($T_1$) curves for full integrated spectrum in (a) before and after annealing at $T_A$ = 250 K.

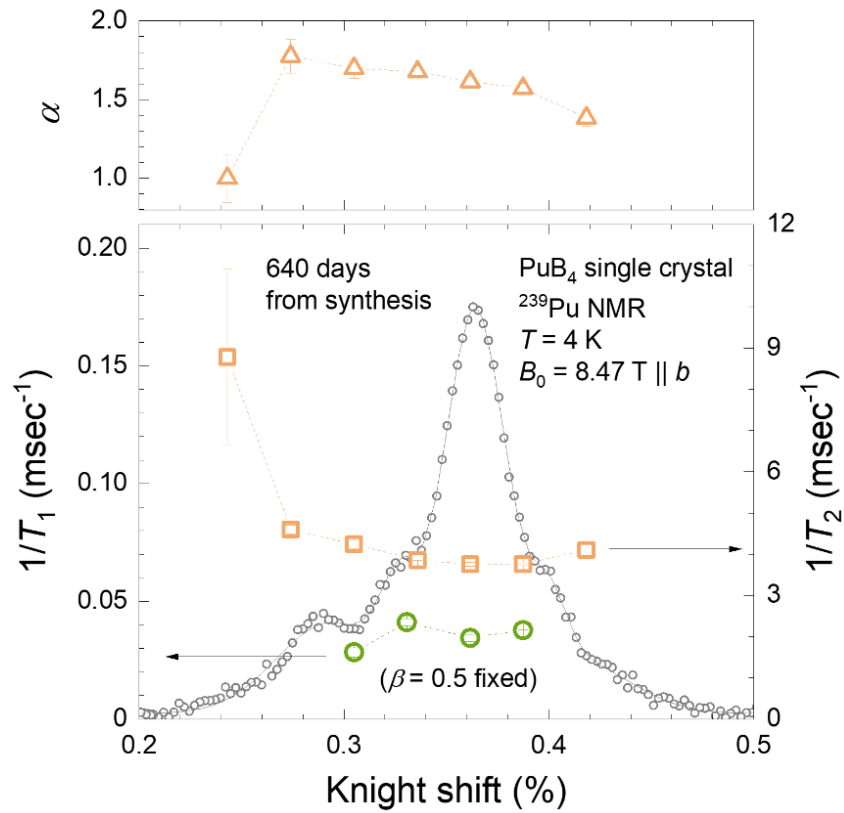

**FIG. 9**. Variation of relaxation parameters in PuB$_4$ single crystal across $^{239}$Pu-NMR spectrum after 640 days from synthesis. Circles represent the spin-lattice relaxation rate $1/T_1$ (left axis), and squares gives the spin-spin relaxation rate $1/T_2$ (right axis), both of which are recorded at 8.47 T for $B_0 \| b$ and are determined from the integrated spectral intensity of a small fraction of the full spectrum. The values of $1/T_1$ were obtained using a fixed stretching exponent of $\beta = 0.5$. Top figure shows the stretching exponent $\alpha$ of the spin echo decay curve throughout the spectrum [see Eq. (2)].

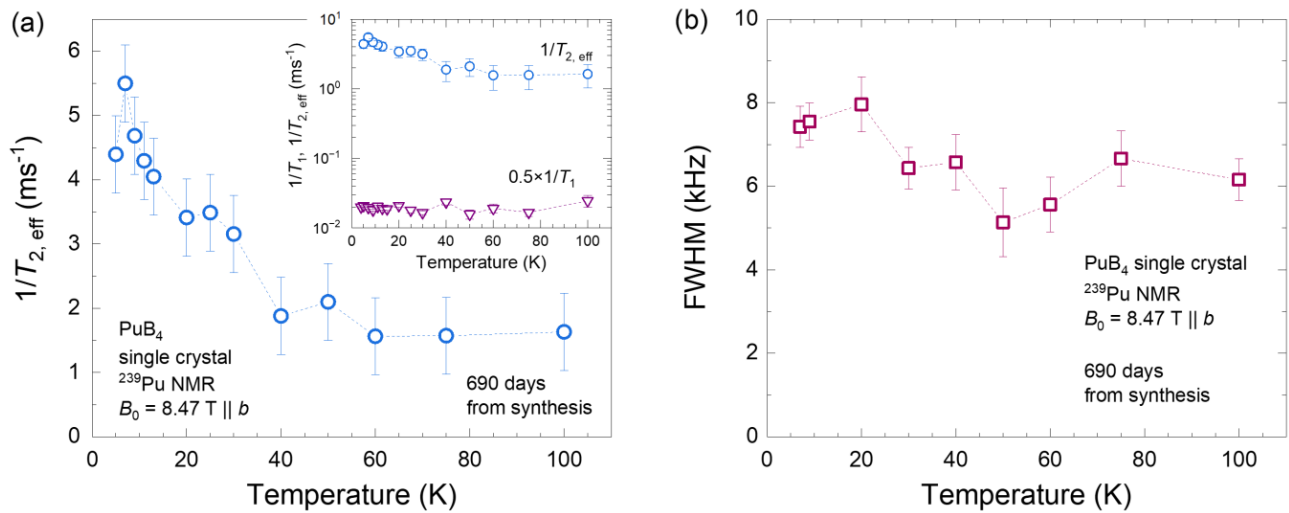

**FIG. 10.** Temperature dependence of $^{239}$Pu-NMR spin-spin relaxation rate $1/T_{2,\,\text{eff}}$ (a) and full width at half maximum (FWHM) (b) in PuB$_4$ single crystal at 690 days from synthesis at 8.47 T for $B_0 \| b$. The values of $1/T_{2,\,\text{eff}}$ are determined in a non-standard way from the spin-echo decay data at 4.2 K (see Appendix A for details). Dashed lines are guides to the eyes. The FWHM is determined from the linewidth of the central peak [the line M in Fig. 3(a)] by performing multi Gaussian fits to the full spectrum. Inset of (a): Temperature dependence of $1/T_{2,\,\text{eff}}$ plotted with the corresponding Redfield contribution $(1/T_1)/2$ [replotted from Fig. 5(a)].

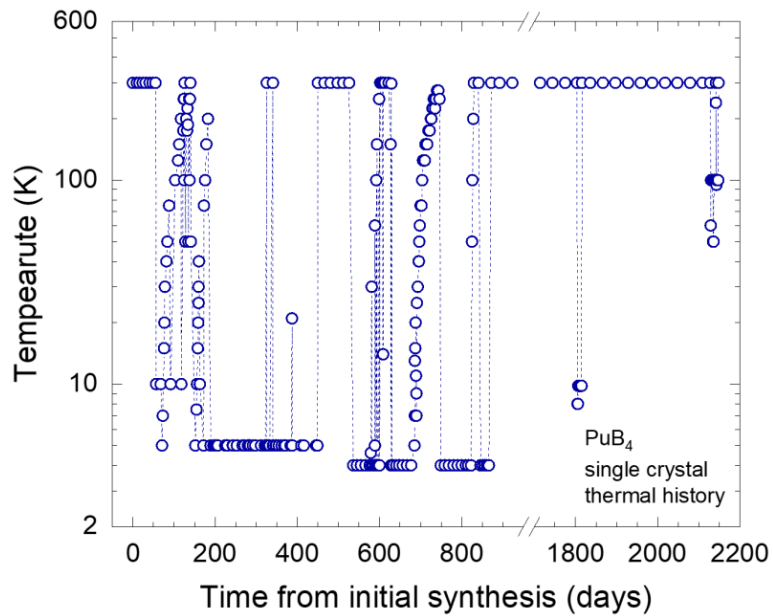

**FIG. S1.** Thermal history of the PuB$_4$ single crystal used in this study. The $^{239}$Pu-NMR experiments at low temperature happened occasionally over in total six years, although the $^{239}$Pu signals were found only within the first three years. Between successive measurements, the sample was stored at room temperature in air.

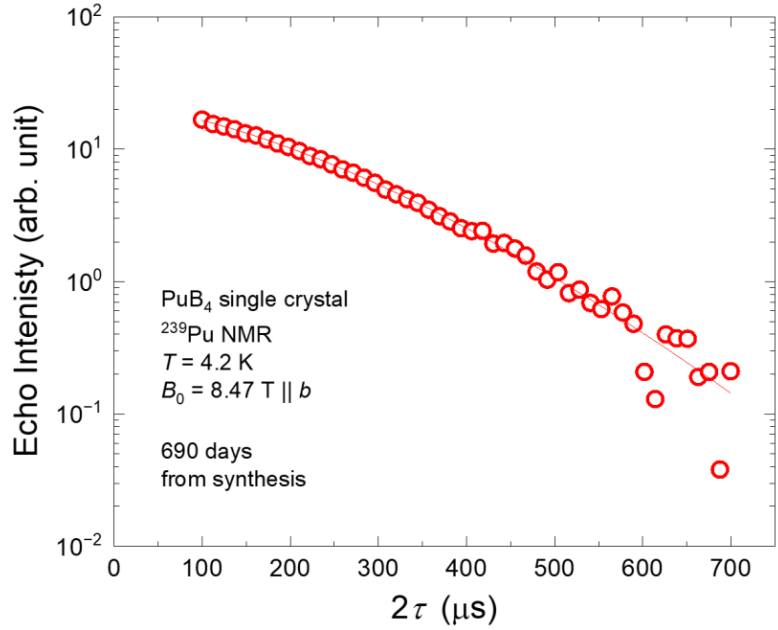

**FIG. S2.** $^{239}$Pu-NMR spin echo decay curve at 4.2 K in PuB$_4$ single crystal at 690 days from synthesis at the external magnetic field $B_0$ of 8.47 T for $B_0 \| b$. The echo intensity corresponding to the full integrated spectral intensity is plotted as a function of the waiting time $\tau$ between $t_{90°}$ and $t_{180°}$ pulses in the spin echo. Solid curve gives the best fit to the data with the form $M_0 \exp[-(2\tau/T_2)^\alpha]$, which yields the spin-spin relaxation time of $T_2 = 243.4$ μs and the stretching exponent of $\alpha = 1.52$.

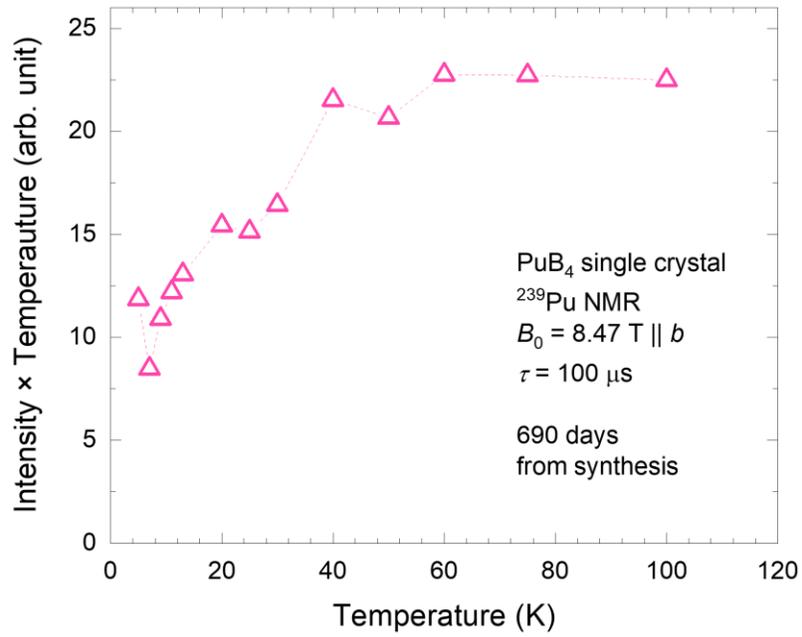

**FIG. S3.** $^{239}$Pu-NMR intensity times temperature plotted against temperature. Data is taken in PuB$_4$ single crystal at 690 days from synthesis at the external magnetic field $B_0$ of 8.47 T for $B_0 \| b$. The waiting time of $\tau = 100$ μs is used in the spin echo measurement.